\documentclass[preprint,a4paper,nofootinbib]{revtex4-1}
\usepackage[utf8]{inputenc}
\usepackage{textcomp}
\usepackage{gensymb}
\usepackage{amsmath} 
\usepackage{amssymb}
\usepackage{xspace}
\usepackage{nicefrac}
\usepackage{xcolor}
\usepackage{graphicx}
\usepackage{setspace}

\newcommand{\dif}{\ifmmode {\mathrm{d}} \fi}
\newcommand{\qgsjet}{{\scshape QGSJetII.04}~}
\newcommand{\urqmd}{{\scshape urqmd}~}
\newcommand{\gcm}{\ifmmode{\mathrm{g/cm}^2}\else{{g/cm}$^2$}\fi}
\newcommand{\xmax}{\ifmmode{X_\mathrm{max}}\else{$X_\mathrm{max}$}\fi \xspace}
\def \deg{\ifmmode{^{\circ}}\else{$^{\circ}$}\fi\xspace} 

\begin{document}

\title{Parametrization of the angular distribution of Cherenkov light in air showers}
\author{Luan B. Arbeletche}
\email{luan.arbeletche@ifsc.usp.br}
\author{Vitor de Souza}
\email{vitor@ifsc.usp.br}
\affiliation{Instituto de F\'isica de S\~ao Carlos, Universidade de S\~ao Paulo, Av. Trabalhador S\~ao-carlense 400, S\~ao Carlos, Brasil.}

\begin{abstract}
    The Cherenkov light produced in air showers largely contributes to the signal observed in ground-based gamma-ray and cosmic-ray observatories. Yet, no description of this phenomenon is available covering both regions of small and large angles to the shower axis. To fill this gap, a parametrization of the angular distribution of Cherenkov photons is performed in terms of a physically-motivated parametric function. Model parameters are constrained using simulated gamma-ray and proton showers with energies in the TeV to EeV region. As a result, a new parametrization is obtained that improves the precision of previous works. Results presented here can be used in the reconstruction of showers with imaging Cherenkov telescopes as well as in the reconstruction of shower profiles with fluorescence detectors.
\end{abstract}

\maketitle

\section{Introduction}
\label{sec:intro}

A large amount of Cherenkov light is produced in extensive air showers~\cite{galbraith_light_1953} and several experimental techniques have been proposed to explore this signal to study astroparticle physics. The generation of light in the cascade is highly dominated by electrons. The emission of Cherenkov light by relativistic electrons including geometry, intensity, and wavelength is explained by classical electrodynamics~\cite{PhysRev.52.378}, which has been used as an inspiration for the development of robust detection techniques.

The total signal produced by all particles in the air shower evolves as the cascade deepens in the atmosphere. The correct description of this evolution is mandatory to extract physical results from measurements. This problem is common to all collaborations running ground-based detectors, including Imaging Atmospheric Cherenkov Telescopes (IACT) and Fluorescence Detectors (FD), and also to proposed space experiments. In particular, to reconstruct the properties of the primary particle, it is necessary to understand the properties of Cherenkov-light production in air showers, including the longitudinal distribution, the lateral distribution, and the angular distribution. In this paper, special attention is given to the description of the angular distribution of Cherenkov photons in air showers.

IACTs are of fundamental importance for the Very High Energy (VHE) gamma-ray astronomy ($E_0 > 100\,$GeV). The identification and the reconstruction of the primary gamma-ray are done by interpreting the Cherenkov light detected by telescopes at ground. Current observatories~\cite{bib:hess,bib:magic,bib:veritas} are equipped with some ($\leq \,$5) telescopes with few degrees ($< \,$5\deg) of field of view installed hundred meters apart from each other. The Cherenkov Telescope Array (CTA)~\cite{bib:cta} is the next-generation IACT system presently under development. The CTA baseline design calls for 118 telescopes to be installed at two sites covering areas of 0.6$\,$km$^2$ in La Palma, Spain and 4$\,$km$^2$ in Paranal, Chile. The angular distribution of Cherenkov photons in an air shower determines the image shape detected by IACTs and is therefore a key aspect in many reconstruction techniques~\cite{bib:3d:rec,bib:performance:hess,bib:performance:magic,bib:performance:hess}.

FDs have been long used to study Ultra-High Energy Cosmic Rays (UHERC)~\cite{bib:flys:eye}. These telescopes have been optimized to measure the isotropic fluorescence light emitted by nitrogen molecules due to the passage of charged particles in the atmosphere. The telescopes in operation~\cite{bib:auger:fd,bib:ta:xmax} have large aperture ($\approx\,$30\deg) and cover a detection area of thousands km$^2$. The fluorescence and Cherenkov emission produce signals in the telescopes in the overlapping wavelength band of 300-450$\,$nm, making it impossible to separate their signals. Traditionally, Cherenkov light was considered as an unwanted noise in the FD measurements~\cite{bib:ta:xmax}, but recently the Cherenkov light seen by FDs has been used as signal to detect showers with energies down to 2$\,$PeV~\cite{Unger:2008:reco,Novotny:spectrum,bib:ta:tale}. Direct Cherenkov light is also used to study UHECR with ground detectors~\cite{budnev_tunka-25_2013,Ivanov_2009} and is proposed to be used as an important signal source in future space experiments~\cite{bib:poemma}. The angular distribution of Cherenkov photons in an air shower is an important feature for all UHECR experiments because it determines the lateral spread of light and the balance between fluorescence and Cherenkov-light signals measured by the FDs, including large angles ($>\,$10\deg) and great distances (several km) from the shower axis.

The number of Cherenkov photons produced in an air shower reaching a detector at a given distance from the shower axis can be calculated only if the angular distribution of photons is known. Reversely, the reconstruction of the primary particle properties is only possible if the measured amount of light in each detector is converted into the amount of light emitted by the particles in the shower. The angular distribution of Cherenkov photons is determined by the convolution of the longitudinal development of
electrons\footnote{The term \textit{electrons} here refer to both electrons and positrons.},
the energy distribution of electrons, the angular distribution of electrons, the scattering of electrons, the refractive index, the geomagnetic field effects, and the scattering of photons~\cite{bib:stanev,Hillas_1982,bib:elbert,Giller200497,NERLING2006421}.

Influenced by the main techniques detecting Cherenkov light (IACT and FD), the study of the angular distribution of Cherenkov photons has been divided respectively in two regimes: (a) gamma-ray primaries, small angles $<$ 10\deg, and TeV energies and (b) cosmic ray primaries, large angles $>$ 10\deg, and highest energies ($10^{17}\,$eV). Experiments have measured the angular distribution of Cherenkov photons~\cite{Baltrusaitis_1987} in regime (b). Since the pioneering work~\cite{Hillas_1982}, the angular distribution was simulated for regime (a)~\cite{bib:3d:rec} and (b)~\cite{NERLING2006421,Giller:2009zz}.

In this paper, the angular distribution of Cherenkov photons is simulated using the most updated simulation software and a new parametrization based on shower physics is proposed. The parametrization presented here improves the precision in the description of the angular distribution of Cherenkov light in comparison to models found in previous publications~\cite{bib:3d:rec,NERLING2006421,Giller:2009zz}. Beside the needed update of the parametrizations concerning the new shower models made available after the previous works, this paper aims at the improvement of the precision requested by the new generation of experiments~\cite{bib:cta,bib:poemma} and at the refinement demanded by the new uses of Cherenkov light as the main signal source in FD analyses~\cite{Novotny:spectrum,bib:ta:tale}. Moreover, a unified view of the two regimes is presented for the first time.

This paper is organized as follows. In section \ref{sec:model:exa}, an exact model to compute the angular distribution of Cherenkov photons is derived. This model is simplified in section \ref{sec:model:app} to obtain a simple form in terms of free parameters. The parameters of the model are constrained by Monte Carlo simulations in section \ref{sec:parametrization}. A discussion of the results and a comparison to previous works are presented in section \ref{sec:results} and some final remarks are given in section \ref{sec:conclusion}.

\section{Exact model for the Cherenkov light angular distribution}
\label{sec:model:exa}

A mathematical description of the number of Cherenkov photons emitted in a given angular interval as a function of the shower development in the atmosphere, $\text{d}^2  N_\gamma / \text{d} \theta \, \text{d} X$, is presented in this section. Each physical quantity relevant to this description is identified and explained below.

\begin{figure}
  \centering
  \includegraphics[width=8cm]{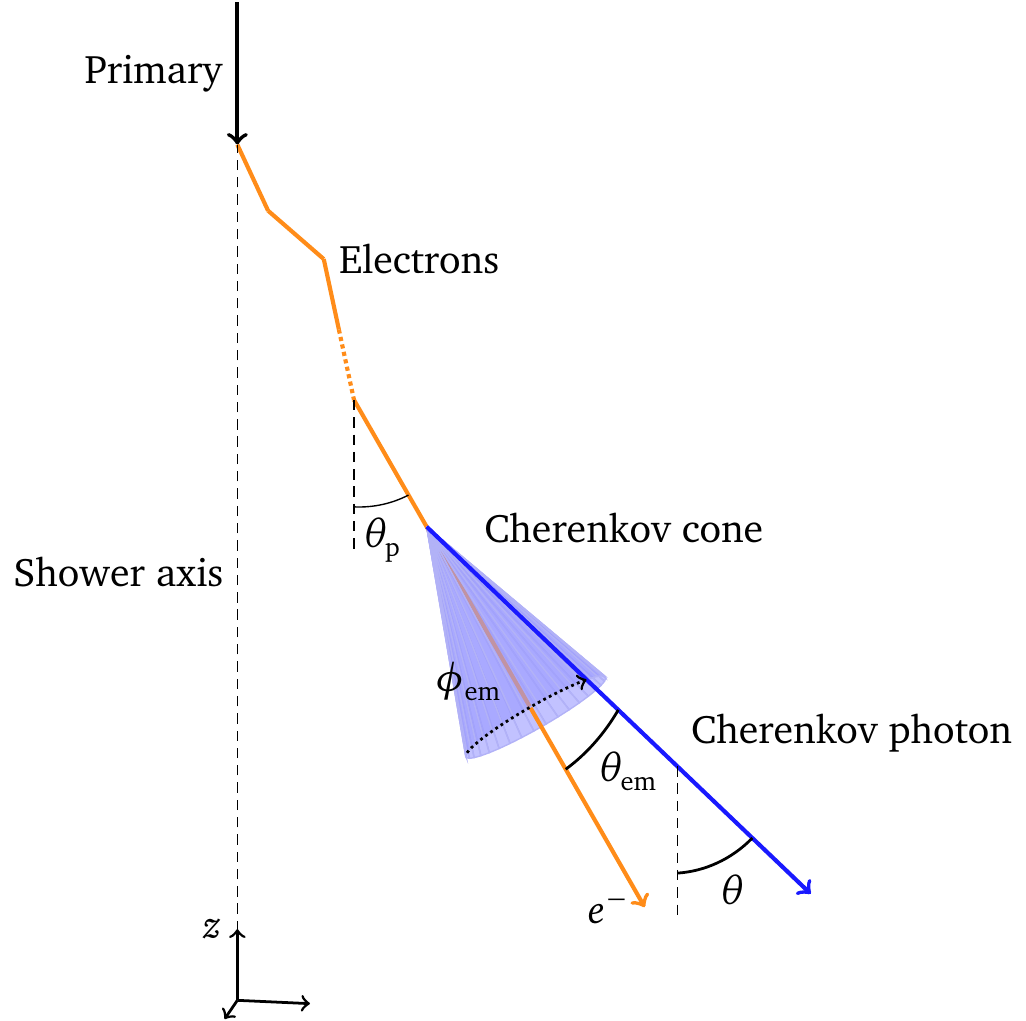}
  \caption{Definition of the relevant angles for Cherenkov light emission in air showers. The blue cone represents the emission of Cherenkov photons around the emitting electron trajectory, in orange. The final angle between each Cherenkov photon (blue trajectory) and the shower axis is denoted by $\theta$.}
  \label{fig:angles}
\end{figure}

Electrons are responsible for over 98\% of the Cherenkov-photon content in a shower~\cite{NERLING2006421}. Therefore it is assumed in this study that all photons are emitted by electrons. Figure \ref{fig:angles} depicts the composition of angles determining the final angular distribution of Cherenkov photons. Shown is that an electron emitted during the shower development is subject to scattering in the atmosphere and its trajectory forms an angle $\theta_\text{p}$ with the shower axis. Such an electron will emit Cherenkov photons in a cone of half-aperture angle $\theta_\text{em}$ around its propagation path. The two angles $\theta_\text{em}$ and $\phi_\text{em}$ (measured in a plane perpendicular to the moving electron track) determine the direction of the emitted photon. Finally, the emitted photon forms an angle $\theta$ with the shower axis.

It is the interplay between the Cherenkov emission angle $\theta_\text{em}$ and the scattering angle of electrons $\theta_\text{p}$ that determines the distribution of the resulting Cherenkov-photon angle $\theta$. In the beginning of the shower, most electrons move parallel to the shower axis, therefore $\theta_\text{p} \approx 0 \Rightarrow \theta \approx \theta_\text{em}$.
%
The angle $\theta_\text{em}$, in its turn, is an increasing function of the atmospheric depth and reaches a maximum value of about $1.5\deg$ at sea level.
As the cascade develops further, electrons scatter multiple times, increasing the fraction of particles with large $\theta_\text{p}$ values. 
%
Indeed, the effect of multiple scattering generates electrons with $\theta_\text{p} > \theta_\text{em}$ and therefore $\theta > \theta_\text{em}$. 
%
As a consequence, for $\theta_\text{p} \gg \theta_\text{em} \Rightarrow \theta \approx \theta_\text{p}$, so that the angular distribution of Cherenkov photons approximately reproduces the angular distribution of electrons in the shower.

The number of Cherenkov photons emitted by electrons with energy $E$ and angle $\theta_\text{p}$ in a shower per interval of depth $\text{d}X$ is given
by\footnote{\setstretch{0.85}The dependency on the primary particle energy ($E_0$) is omitted here for brevity and discussed in terms of simulated showers in the following sections. For the purpose of this section, $E_0$ may be regarded as fixed.}
\begin{equation}
    \text{d}N_\gamma
    =
    N_e(s) \,
    \frac{\text{d}N_e}{\text{d}E}(E,s) \, \text{d}E \,
    \frac{\text{d}N_e}{\text{d}\theta_\text{p}}(\theta_\text{p},E) \, \text{d}\theta_\text{p} \,
    Y_\gamma(E,h) \, \sec \theta_\text{p} \, \text{d}X \, 
    \frac{\text{d}\phi_\text{em}}{2\pi} \, ,
    \label{eq:differential1}
\end{equation}
where $s$ is the shower
age\footnote{$s = 3X/(X + 2X_\text{max})$ where $X_\text{max}$ is the depth in which the shower reaches the maximum number of particles.}
and $h$ is the emission height above sea level. $N_e(s)$ is the total number of electrons, $\text{d}N_e/\text{d}E$ is the energy distribution of electrons, and $\text{d}N_e/\text{d}\theta_\text{p}$ is the angular distribution of electrons. The function $Y_\gamma(E,h)$ represents the number of photons emitted by one electron per depth interval (yield) and the factor of $\sec\theta_\text{p}$ takes into account the correction in the length of the electron track due to its inclined trajectory. Photons are uniformly distributed in  $\phi_\text{em}$ (factor of $1/2\pi$). According to reference~\cite{NERLING2006421}, $Y_\gamma(E,h)$ is given by
\begin{equation}
\begin{split}
    Y_\gamma(E,h)
    &\approx
    4\pi \, \alpha \,
    \frac{ n(h)-1}{\rho(h)}
    \left( \frac{1}{\lambda_1} - \frac{1}{\lambda_2}\right)
    \left( 1 - \frac{E^2_\mathrm{thr}(h)}{E^2} \right) \, ,
    \label{eq:cer-yield}
\end{split}
\end{equation}
in which $\alpha \approx \nicefrac{1}{137}$ is the fine-structure constant, $n(h)$ is the refractive index of the medium, $\rho(h)$ is the atmospheric density, and $\lambda_i$ the wavelength interval of the emitted photons. The threshold energy $E_\mathrm{thr}$ for an electron to  produce Cherenkov light is $E_\mathrm{thr}(h) = m_{e}c^2/\sqrt{1-n^{-2}(h)}$, where $m_e$ is the electron rest mass.

\begin{figure}
  \centering
  \includegraphics[width=8cm]{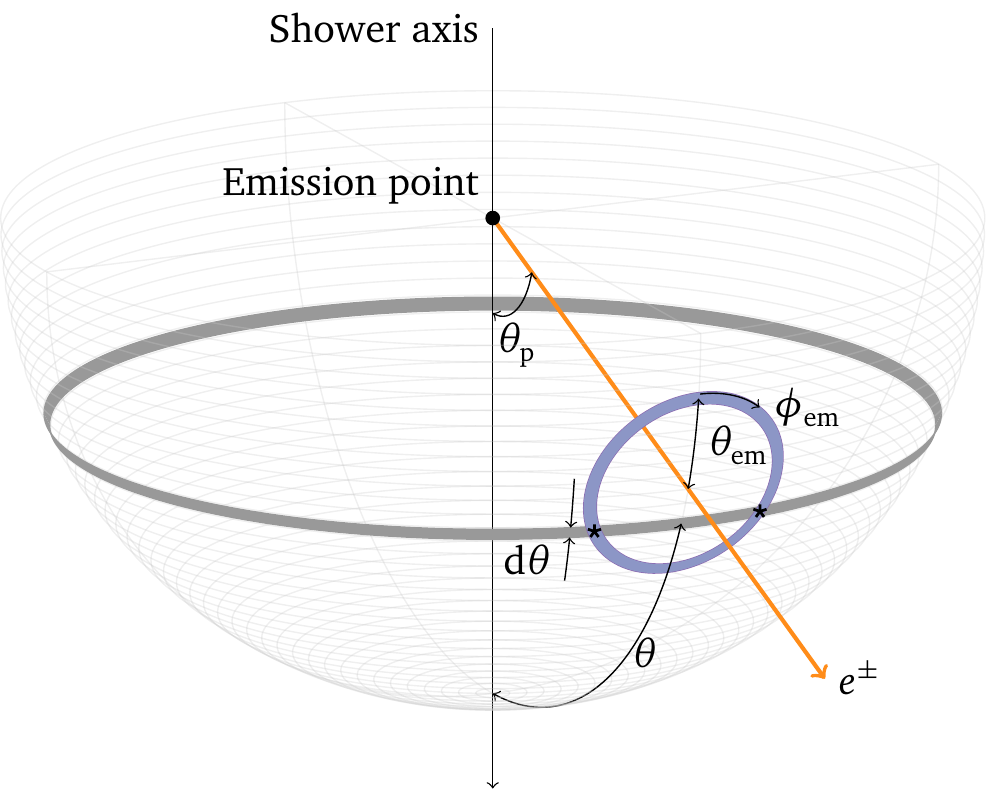}
  \caption{Depiction of the intersecting region between the Cherenkov cone (blue ring) and the ring of width $\text{d}\theta$ around the angle $\theta$ (grey ring) in the unit sphere. There are two intersection points whenever $|\theta-\theta_\text{p}| < \theta_\text{em}$ and none otherwise.}
  \label{fig:sphere}
\end{figure}

The dependency of $\text{d}N_\gamma$ on the angle between the Cherenkov photon and the shower axis directions, $\theta$, is found after a change of variable from $\phi_\text{em}$ to $\theta$ (see Figure \ref{fig:angles})
\begin{equation}
    \cos\theta
    =
    \cos\theta_\text{p} \cos\theta_\text{em}
    -
    \sin\theta_\text{p} \sin\theta_\text{em} \cos\phi_\text{em} \, ,
    \label{eq:angle}
\end{equation}
which leads to
\begin{equation}
\begin{split}
    \text{d}\phi_\text{em}
    &=
    2 \left| \frac{\text{d}\phi_\text{em}}{\text{d}\theta} \right| \text{d}\theta 
    \\
    &=
    \frac{2\; \sin\theta\;\text{d}\theta}{\sqrt{\sin^2\theta_\text{p} \sin^2\theta_\text{em} - (\cos \theta_\text{p} \cos \theta_\text{em} -\cos \theta)^2}} \, ,
    \label{eq:change-of-variable}
\end{split}
\end{equation}
in which a factor of $2$ was added to account for the fact that there are always two values of $\phi_\text{em}$ resulting in the same value of $\theta$ (see Figure \ref{fig:sphere}). The half-aperture angle of the Cherenkov radiation cone, $\theta_\text{em}$, relates to the particle velocity $\beta$ by the usual expression 
\begin{equation}
    \cos\theta_\text{em} = \frac{1}{\beta\, n} \; .
    \label{eq:cone-angle}
\end{equation}

Substitution of Equation \eqref{eq:change-of-variable} into Equation \eqref{eq:differential1} gives
\begin{equation}
\begin{split}
    \text{d}N_\gamma
    =
    & \, N_e(s) \, 
    \frac{\text{d}N_e}{\text{d}E}(E,s) \, \text{d}E \,
    \frac{\text{d}N_e}{\text{d}\theta_\text{p}}(\theta_\text{p},E) \, \text{d}\theta_\text{p} \,
    Y_\gamma(E,h) \, \sec \theta_\text{p} \, \text{d}X
    \\
    & \times \,
    \frac{1}{\pi}
    \frac{\sin\theta \, \text{d}\theta}{\sqrt{\sin^2\theta_\text{p} \sin^2\theta_\text{em} - (\cos \theta_\text{p} \cos \theta_\text{em} -\cos \theta)^2}} \, .
    \label{eq:differential2}
\end{split}
\end{equation}

Finally, to obtain the desired angular distribution of Cherenkov photons, $\text{d}^2  N_\gamma / \text{d} \theta \, \text{d} X$, it is necessary to integrate Equation \eqref{eq:differential2} over all possible values of electron energies $E$ and angles $\theta_\text{p}$. Integration over $E$ must assert that relation \eqref{eq:cone-angle} is satisfied, therefore $E$ takes values for which $E > E_\text{thr}(h)$. Limits of the integral over electron angles $\theta_\text{p}$ should take only values that contribute to $\theta$. From Figure \ref{fig:sphere} and Equation \eqref{eq:angle}, it is found that this interval is $|\theta-\theta_\text{em}| < \theta_\text{p} < \theta+\theta_\text{em}$. Thus, the exact angular distribution of Cherenkov photons is given by
\begin{equation}
\begin{split}
  \frac{\text{d}^2N_\gamma }{\text{d}\theta \, \text{d}X} (\theta,s,h)
  = 
  & \, \frac{1}{\pi} \, N_\text{e}(s) \, \sin \theta 
  \int_{E_\mathrm{thr}(h)}^\infty \text{d}E  \, Y_\gamma(E,h) \, \frac{\text{d}N_\text{e}}{\text{d}E}(E,s) \\
  & \times
  \int_{|\theta - \theta_\text{em}|}^{\theta+\theta_\text{em}} \frac{\text{d}N_e}{\text{d}\theta_\text{p}}(\theta_\text{p},E) \, \frac{\text{d}\theta_\text{p}}{\cos\theta_\mathrm{p} \sqrt{\sin^2 \theta_\text{p} \sin^2 \theta_\text{em} - (\cos \theta_\text{p} \cos \theta_\text{em} - \cos \theta)^2}} \, .
\end{split}
\label{eq:dedthetadx}
\end{equation}

\section{Approximated model for the Cherenkov light angular distribution}
\label{sec:model:app}

In this section, an approximation of the above equation is proposed to obtain a simpler yet meaningful description of the angular distributions of Cherenkov light. The idea is to summarize the angular distribution to a minimum set of parameters, allowing its parametrization.

First, note that the integration in $\theta_\text{p}$ is done in a very narrow interval given that $\theta_\mathrm{em}<1.5\degree$. Therefore it is possible to consider that $\sec\theta_\text{p} \, \text{d}N_e / \text{d}\theta_\text{p}$ varies little within integration limits and, in a first approximation, can be taken as constant and calculated in the mean angle $\langle\theta_\text{p}\rangle$ of the range in between the integration limits
\begin{equation}
\begin{split}
  \frac{\text{d}^2N_\gamma }{\text{d}\theta \, \text{d}X} (\theta,s,h)
   \approx 
   & \, \frac{1}{\pi} \, N_\text{e}(s) \, \sin \theta \int_{E_\mathrm{thr}(h)}^\infty \text{d}E  \; Y_\gamma(E,h) \, \frac{\text{d}N_\text{e}}{\text{d}E}(E,s) \\
   & \times \frac{1}{\cos{\langle\theta_\text{p}\rangle}} \frac{\text{d}N_e}{\text{d}\theta_\text{p}}({\langle\theta_\text{p}\rangle},E) \\
   &\times \int_{|\theta - \theta_\text{em}|}^{\theta+\theta_\text{em}}\, \frac{\text{d}\theta_\text{p}}{ \sqrt{\sin^2 \theta_\text{p} \sin^2 \theta_\text{em} - (\cos \theta_\text{p} \cos \theta_\text{em} - \cos \theta)^2}} \, ,
\end{split}
\label{eq:app:1}
\end{equation}
where
\begin{equation}
  \langle\theta_\text{p}\rangle =
  \begin{cases}
    \theta_\text{em} \,, \; \; &\text{if}  \; \; \; \theta \leq \theta_\text{em} \\
    \theta  \,, \; \; &\text{if}  \; \; \; \theta > \theta_\text{em}
  \end{cases}
  \, .
\end{equation}

The remaining integral over $\theta_\text{p}$ is a complete elliptic integral of the first kind and can be approximated by a logarithmic function
\begin{equation}
    \begin{split}
    \int_{|\theta - \theta_\text{em}|}^{\theta+\theta_\text{em}} \, \frac
    {\text{d}\theta_\text{p}}
    { \sqrt{\sin^2 \theta_\text{p} \sin^2 \theta_\text{em} - (\cos \theta_\text{p} \cos \theta_\text{em} - \cos \theta)^2}}
    \approx 
    \\
    \frac{1}{\sin\langle\theta_\text{p}\rangle}
    \begin{cases}
       \pi - \log\big(1 - \frac{\theta}{\theta_\mathrm{em}}\big) 
       \;, \; \; &\text{if}  \; \; \; \theta \leq \theta_\mathrm{em}
        \\
        \pi - \log\big(1 - \frac{\theta_\mathrm{em}}{\theta}\big)
        \;, \; \; &\text{if}  \; \; \; \theta > \theta_\mathrm{em}
    \end{cases}
    \, .
    \end{split}
\end{equation}
The abbreviation below is introduced
\begin{equation}
    I(\theta,\theta_\mathrm{em},E)
    =
    \frac{1}{\sin\langle\theta_\text{p}\rangle}
    \begin{cases}
       \pi - \log\big(1 - \frac{\theta}{\theta_\mathrm{em}}\big) 
       \;, \; \; &\text{if}  \; \; \; \theta \leq \theta_\mathrm{em}
        \\
        \pi - \log\big(1 - \frac{\theta_\mathrm{em}}{\theta}\big)
        \;, \; \; &\text{if}  \; \; \; \theta > \theta_\mathrm{em}
    \end{cases}
    \, ,
    \label{eq:I:def}
\end{equation}
and by noting that $\cos\theta_\text{em} = 1/\beta n$ rapidly converges to $1/n$ as the electron energy increases, it is reasonable to assume that $\cos\theta_\text{em} = 1/n$ for all electrons. With this assumption the function $I(\theta,\theta_\text{em},E)\sim I(\theta,\theta_\text{em}) =I(\theta,h)$ becomes independent of the electron energy\footnote{From now on $\theta_\text{em} = \arccos(1/n)$.}
\begin{equation}
\begin{split}
  \frac{\text{d}^2N_\gamma }{\text{d}\theta \, \text{d}X} (\theta,s,h)
  \approx 
  & \, \frac{1}{\pi} \, N_\text{e}(s) \,  \sin\theta \, I(\theta,h) \\ 
  & \times \int_{E_\mathrm{thr}(h)}^\infty \text{d}E  \; Y_\gamma(E,h) \; \frac{\text{d}N_\text{e}}{\text{d}E}(E,s) \; \frac{1}{\cos \langle \theta_\text{p} \rangle}\frac{\text{d}N_e}{\text{d}\theta_\text{p}}(\langle \theta_\text{p} \rangle, E) \; .
\end{split}
\label{eq:app:2}
\end{equation}

The validity of the approximations done until here were tested using Monte Carlo simulations of air showers and the results shown in Appendix~\ref{app:tests}.

The remaining integral over electron energies,
\begin{equation}
  \int_{E_\mathrm{thr}(h)}^\infty \text{d}E  \; Y_\gamma(E,h) \; \frac{\text{d}N_\text{e}}{\text{d}E}(E,s) \frac{1}{\cos \langle \theta_\text{p} \rangle}\frac{\text{d}N_e}{\text{d}\theta_\text{p}}(\langle \theta_\text{p} \rangle, E) \; ,
  \label{eq:app:3}
\end{equation}
has been studied before in references~\cite{Giller:2009zz,NERLING2006421}. A parametric form to describe this quantity is proposed here
\begin{equation}
    K(\theta,s,h) = 
    C \, 
    {\langle\theta_\text{p}\rangle}^{\nu - 1} 
    \mathrm{e}^{-{\langle\theta_\text{p}\rangle}/\theta_1} 
    \left( 1 + \epsilon \,  \text{e}^{{\langle\theta_\text{p}\rangle}/\theta_2} \right) \, ,
    \label{eq:k:def}
\end{equation}
where  $\nu$, $\theta_1$, $\theta_2$, and $\epsilon$ are parameters varying with shower age, height (or refractive index), and, possibly, the primary energy. The constant $C$ is intended to normalize Equation~\eqref{eq:k:def} according to Equation~\eqref{eq:app:3}. In the next section the parameters of this function are going to be studied and the quality of the description is going to be tested. The approximated model is summarized as
\begin{equation}
  \frac{\text{d}^2N_\gamma }{\text{d}\theta \; \text{d}X} (\theta,s,h)
   =
   \frac{1}{\pi} \, N_\text{e}(s) \, \sin \theta \, I(\theta,h) \, K(\theta,s,h).
\label{eq:app:final}
\end{equation}

\section{Parametrization of the Cherenkov light angular distribution}
\label{sec:parametrization}

Monte Carlo simulations of air showers are done using the CORSIKA$\,$7.6900 package~\cite{Heck:1998vt}. Gamma-ray and proton showers are simulated with energies between $100\,$GeV and $1\,$EeV in intervals of 1 in $\log_{10}(E_0/\mathrm{eV})$. For each combination of primary type and energy, at least 120 showers are simulated. Simulations are performed for vertical showers and showers inclined at 20\degree. \qgsjet\cite{Ostapchenko:2010vb} and \urqmd\cite{Bleicher:1999xi} are used as high- and low-energy hadronic interaction models, respectively. The U.S. standard atmosphere model is used in the simulations and the refractive index is considered to be independent of the wavelength ($180\,\text{nm} \leq \lambda \leq 700\,\text{nm}$) of the emitted photons. Cherenkov photons are produced in bunches of maximum five. The COAST option is used to store the angle between the Cherenkov photons and the shower axis directions, $\theta$. \xmax, which is used to compute the shower age, is extracted from the longitudinal development of charged particles by fitting a Gaisser-Hillas function~\cite{gaisser-hillas}.

\begin{figure}
    \centering
    \includegraphics{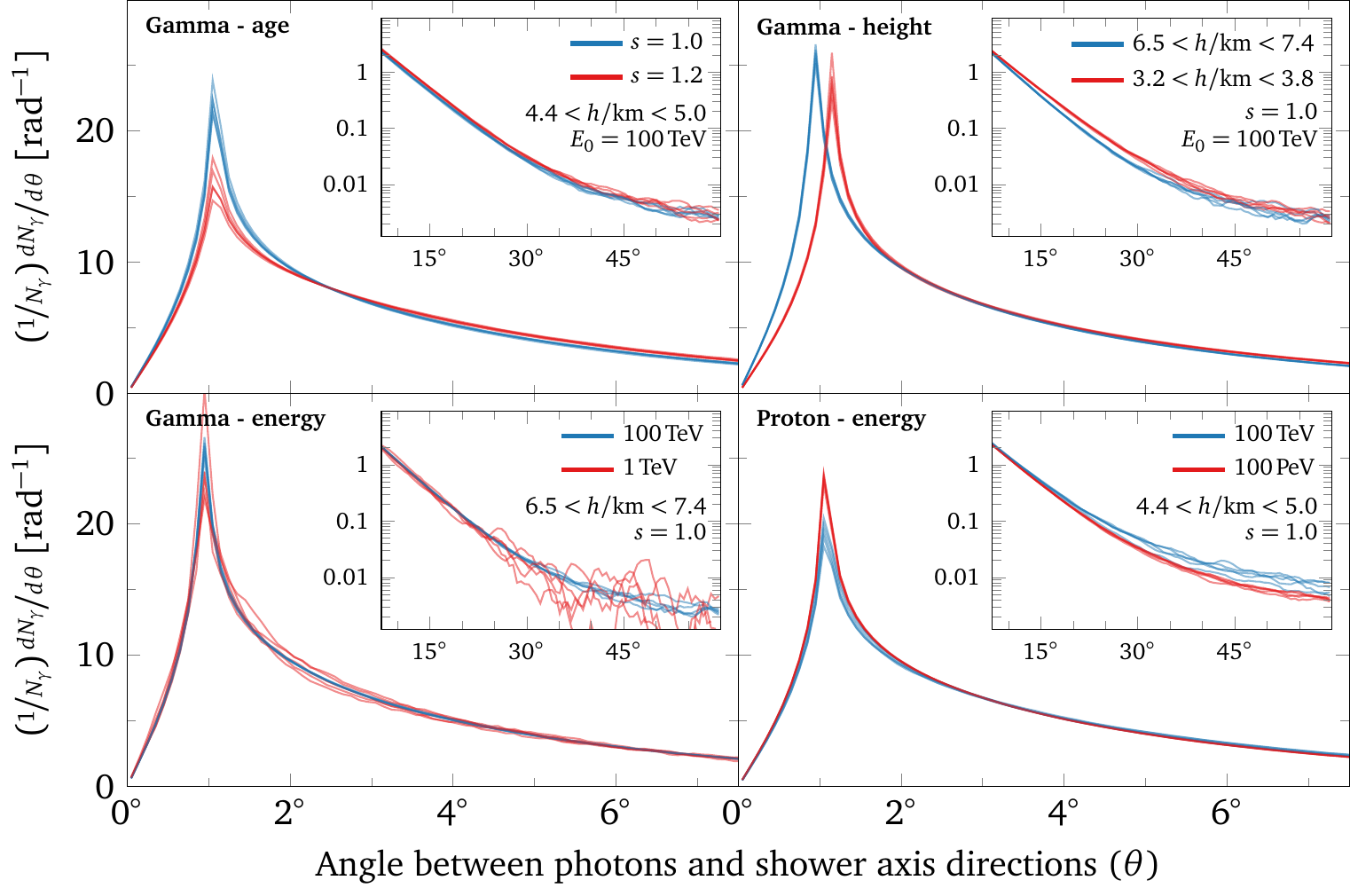}
    \caption{Examples of simulated angular distributions of Cherenkov photons highlighting its dependency with respect to the shower age (top left), emission height (top right), primary energy in gamma-ray showers (bottom left), and primary energy in proton showers (bottom left).}
    \label{fig:sim:dist}
\end{figure}

The approximated model summarized in Equation~\eqref{eq:app:final} suggests that the angular distribution of Cherenkov photons should vary with shower age and atmospheric height. Both dependencies are made clear in the upper plots of Figure~\ref{fig:sim:dist}, where the angular distribution of Cherenkov photons of five randomly chosen gamma-ray-induced air showers at different values of $s$ and $h$ are compared. From the cascade theory~\cite{rossi_cr,kamata_nishimura,greisen_review}, the angular distributions of Cherenkov light in gamma-ray showers are expected to be independent of the primary particle energy and this is confirmed in the bottom-left plot of Figure~\ref{fig:sim:dist}. In the case of proton showers, some dependency on the primary energy is observed in the bottom-right plot of the same figure. These plots also reiterate the fact that distributions with common age, height, primary type, and primary energy are similar.

Taking this dependency into account, the angular distribution of Cherenkov photons in a given interval with mean age $\bar{s}$ and height $\bar{h}$ in a shower of energy $E_0$ can be described by
\begin{equation}
  \frac{\text{d}N_\gamma }{\text{d}\theta}(\theta,\bar{s},\bar{h},E_0)
  = 
  N \, \sin\theta \, I(\theta,\bar{h}) \, K(\theta,\bar{s},\bar{h}, E_0) \, ,
\label{eq:app:fit}
\end{equation}
in which $N$ (different from $N_e(s)$) is a normalization constant which depends on the parameters of $K(\theta, \bar{s},\bar{h}, E_0)$.

The parameters of $K(\theta, \bar{s},\bar{h}, E_0)$ are considered to be
\begin{equation}
\begin{split}
    \nu (s,n)     &= p_{0,\nu} \left(n - 1\right)^{p_{1,\nu}} + p_{2,\nu} \log(s) \, , \\
    \theta_1(s,n,E_0) &= p_{0,\theta_1} \left(n - 1\right)^{p_{1,\theta_1}} \left(E_0/\text{TeV}\right)^{p_{2,\theta_1}} + p_{3,\theta_1} \log(s)\, , \\
    \theta_2(s,n) &= \theta_1(s,n) \, \left( p_{0,\theta_2} + p_{1,\theta_2} s \right) \, , \\
    \epsilon(E_0) &= p_{0,\epsilon} + p_{1,\epsilon} \left(E_0/\text{TeV}\right)^{p_{2,\epsilon}} \; .
\end{split}
\end{equation}
The coefficients $p_{i,\mu}$ are the parameters of the model to be fitted. In these equations, the dependence in height, $h$, was changed by the dependence in the refractive index, $n$, to make the parametrization independent of the atmospheric model used in the simulations.

The simulated angular distributions of Cherenkov photons are fitted with this model. For that, a multinomial likelihood function ($L_\text{MLE}$) is built taking into account every simulated distribution from shower ages in the interval $0.8 \leq s \leq 1.2$. A single value of refractive index $n$ is associated with each distribution according to the emission height. Histograms are weighted by the inverse of the primary energy in TeV, so the contribution from showers of distinct energies to $L_\text{MLE}$ are of the same order of magnitude. Gamma-ray and proton showers are fit separately, as distributions strongly depend on the primary particle type in lower energies. All coefficients $p_{i,\mu}$ are allowed to vary in the fit procedure. In the case of gamma showers, however, the energy dependency is dropped ($p_{2,\theta_1}$, $p_{1,\epsilon}$, $p_{2,\epsilon}=0$). Fitted values of $p_{i,\mu}$ and their associated confidence intervals are found in Tables~\ref{tab:par:gamma} and \ref{tab:par:proton}. 

\section{Results}
\label{sec:results}

\begin{figure}
    \centering
    \includegraphics{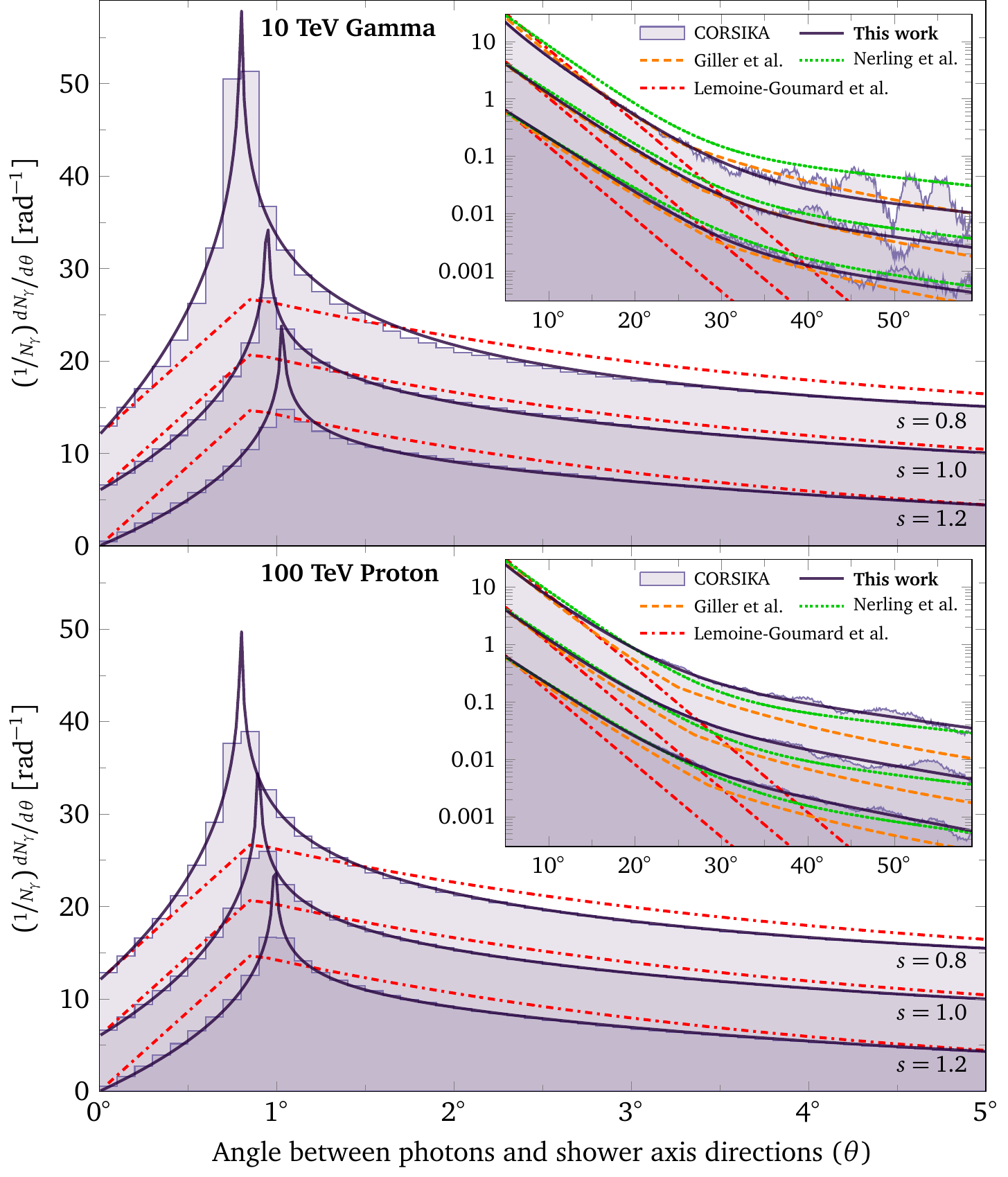}
    \caption{Angular distribution of Cherenkov photons from a single gamma-ray (top) and proton (bottom) shower. CORSIKA simulations (filled histograms) are compared to the parametrized distributions (solid curves) at $s=0.8$, $1.0$, and $1.2$ (indicated inside the plot). Predictions from  Refs. \cite{NERLING2006421,Giller:2009zz,bib:3d:rec} are shown for comparison (see legend). Curves of a common shower age are vertically displaced for better visualization.}
    \label{fig:fit:ex}
\end{figure}

In this section, the parametrization proposed in the previous section is compared to the Monte Carlo distributions and to previous works. Figure~\ref{fig:fit:ex} shows the simulated angular distribution of Cherenkov photons in comparison to four models for one single gamma-ray (upper panel) and one single proton shower (lower panel). The ability of the presented parametrization to describe the simulated data both around the peak of the distributions and at the small and large $\theta$ regions is evident in this figure.
Predictions from the models presented in Refs. \cite{NERLING2006421,Giller:2009zz} are shown in the region $\theta > 5\deg$ only, for which they are defined.

\begin{figure}
    \centering
    \includegraphics{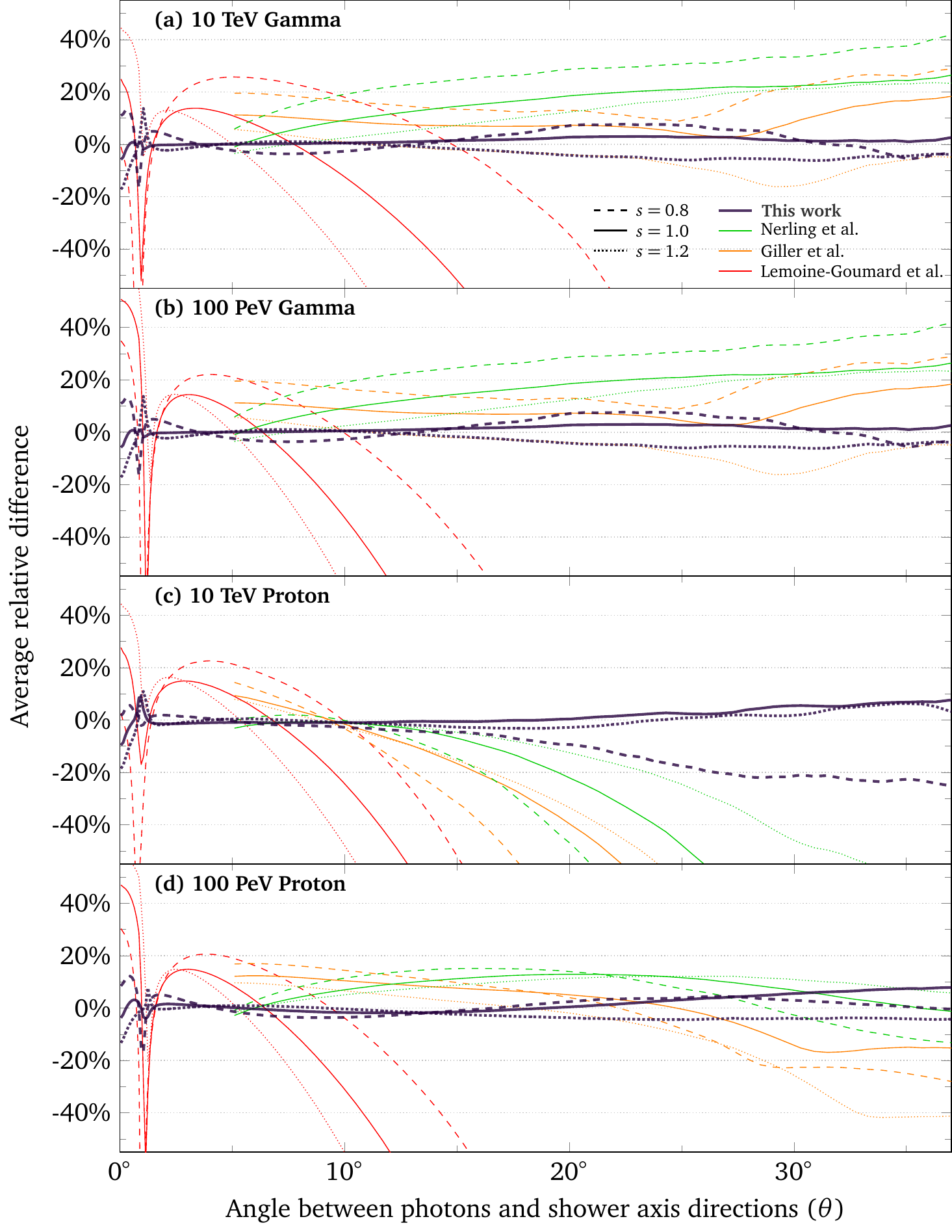}
    \caption{Average relative deviation between parametrized and simulated angular distributions of Cherenkov photons at $s=0.8$ (dashed curves), $1.0$ (solid curves), and $1.2$ (dotted curves). Each box depicts a single primary-energy combination, indicated in the top-left corner. Parametrization of this work is compared to predictions from Refs. \cite{NERLING2006421,Giller:2009zz,bib:3d:rec} (see legend in the upper pannel).}
    \label{fig:fit:dev}
\end{figure}

Figure~\ref{fig:fit:dev} shows the overall quality of the models by comparing their average relative deviation to the simulated distributions for four combinations of primary type and energy. Three shower ages ($s=0.8$, $1.0$, and $1.2$) are shown. It is clear that the model proposed here has many advantages. The model presents the smaller deviation from the simulations for a large angular range. For gamma-ray showers, the deviation is very small ($< 5\%$) for angles smaller than $25^\circ$. For proton showers, the deviation of the model developed here improves with energy.

\begin{figure}
    \centering
    \includegraphics{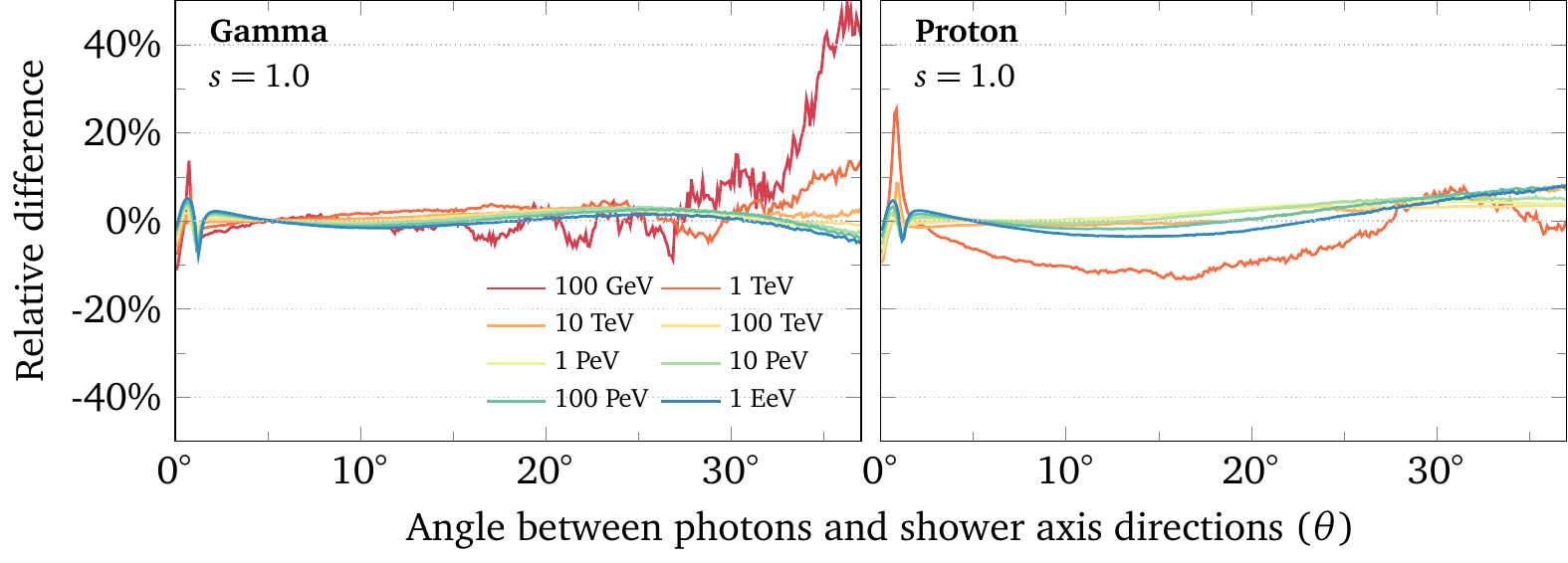}
    \caption{Average relative deviation between parametrized and simulated angular distributions of Cherenkov photons at $s=1.0$ for gamma-ray (left) and proton (right) showers at various primary energies.}
    \label{fig:fit:dev:energy}
\end{figure}

Results presented in this work are optimized with simulated showers having energies from 100$\,$GeV (1$\,$TeV, in case of proton) to 1$\,$EeV. The quality of the model measured with respect to shower energy can be assessed in Figure~\ref{fig:fit:dev:energy}, where the average relative deviation is shown at $s=1.0$ for all studied energies. For both primaries, this deviation is smaller than 10\% in a wide angular and energy interval. In the case of gamma-ray showers (left box), the big deviation seen for 100$\,$GeV gamma-ray showers above $\theta = 30\deg$ is related to the small number of photons in this region. On average, $99.8\%$ of the Cherenkov photon content of 100 GeV gamma-ray fall in the region $\theta < 30\deg$. This figure confirms again the quality of the proposed model and ensures its adequacy to be employed in both the aforementioned regimes (a) and (b).

\section{Conclusion}
\label{sec:conclusion}

To understand the nature and to describe the angular distribution of Cherenkov photons in air showers is of great importance in current experimental astrophysics. An exact model has been derived in Section~\ref{sec:model:exa} to describe the angular distribution photons in terms of the unknown energy and angular distributions of electrons. In Section~\ref{sec:model:app}, successive approximations to this exact model have lead to a factorized form for the angular distributions of Cherenkov photons: a first term, $I$, depending on the maximum Cherenkov emission angle and a second term, $K$, depending on the energy and angular distributions of electrons. A simple parametric form has been proposed to describe this second term, overcoming the necessity of describing the two-dimensional energy and angular distribution of electrons. Parameters of this model have been obtained by fitting Monte Carlo simulations in Section~\ref{sec:parametrization}.

The direct comparison of the parametrization and Monte Carlo simulations in Section~\ref{sec:results} has shown the excellent capability of the model to describe the angular distributions of Cherenkov photons. The use of this model has many advantages as it is able to: 1) cover both small and large angular regions, including the peak around $\theta_\text{em}$ and 2) cover a large energy interval, from hundreds of GeV to EeV energies. The parametrization presented here is therefore adequate to be employed in both the reconstruction of gamma-rays and cosmic-rays in IACT systems and also in the study of extensive air showers with fluorescence detectors.

\section*{Acknowledgments}
LBA and VdS acknowledge FAPESP Project 2015/15897-1.  This study was financed in part by the Coordena\c{c}\~ao de Aperfei\c{c}oamento de Pessoal de N\'{\i}vel Superior - Brasil (CAPES) - Finance Code 001.  Authors acknowledge the National Laboratory for Scientific Computing  (LNCC/MCTI,  Brazil)  for  providing  HPC  resources of the SDumont supercomputer (http://sdumont.lncc.br). VdS acknowledges CNPq.

\appendix
\section{Validation of approximations}
\label{app:tests}

\begin{figure}
  \centering
    \includegraphics{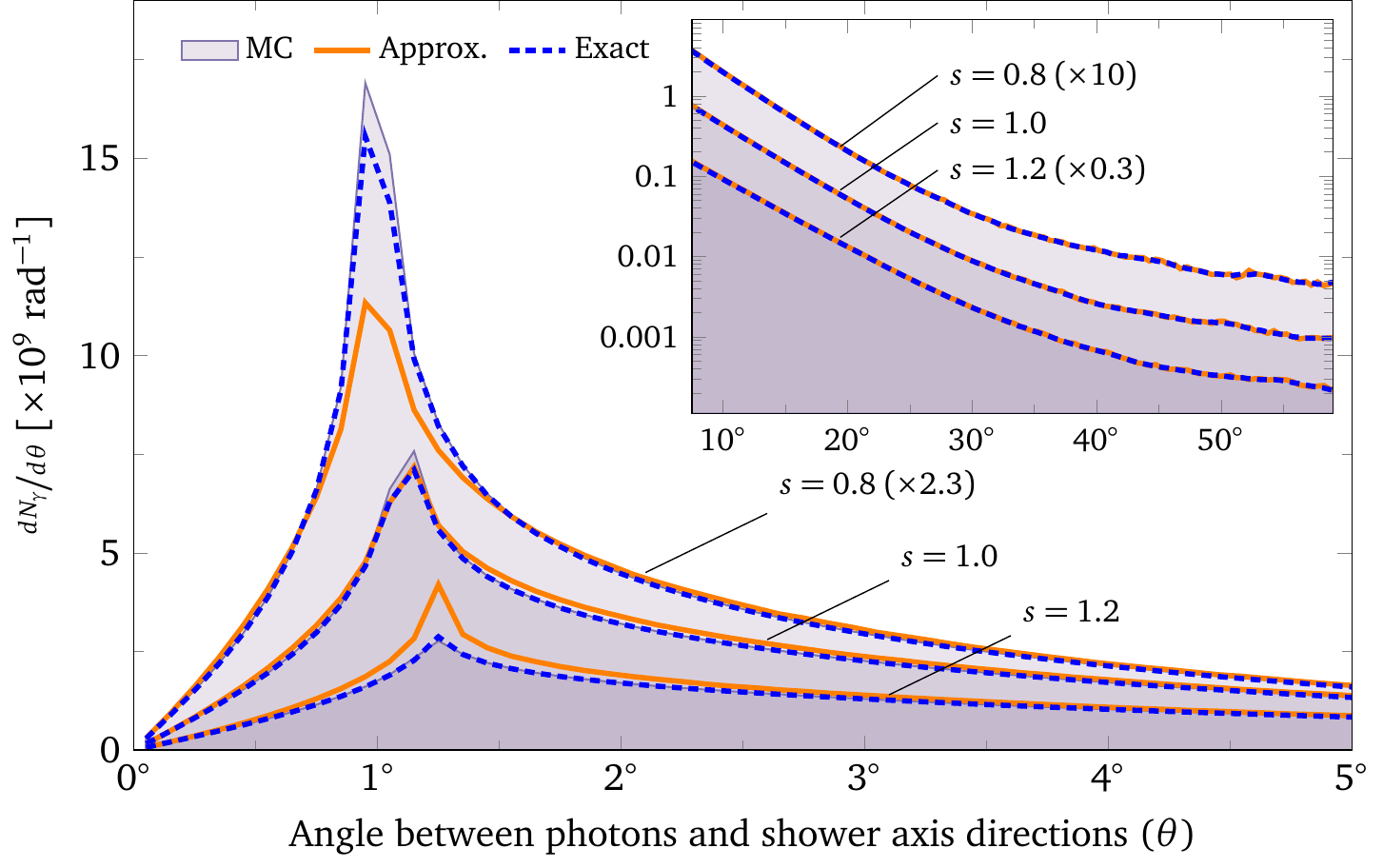}
  \caption{Comparison of the exact (blue dashed) and the approximated (orange solid) models of Sections \ref{sec:model:exa} and \ref{sec:model:app} to a simulated angular distribution of Cherenkov photons at three shower ages (see annotations inside the box). The inset plot shows the region $\theta > 5\degree$. To avoid superposition of the curves and therefore make the figure more clear, some curves were scaled by factors indicated together with the shower age.}
  \label{fig:validation}
\end{figure}

In this Appendix the models presented in Section \ref{sec:model:exa} (exact) and Section \ref{sec:model:app} (approximated) are compared to a direct simulation of the angular distribution of Cherenkov photons. This comparison is shown in Figure \ref{fig:validation} for the case of a vertical $1\,$PeV gamma-ray air shower at three different shower ages. The inset plot shows the region of large angles ($\theta > 5\degree$). The angular distributions of Cherenkov photons directly extracted from the simulation (reference) are represented by the filled curves. The dashed blue (exact model) and solid orange (approximated model) lines show the computation of the angular distributions of Cherenkov photons using Equations \eqref{eq:dedthetadx} and \eqref{eq:app:final}, respectively. For these computations, the energy and angular distributions of electrons ($\text{d}N_e/\text{d}E$ and $\text{d}N_e/\text{d}\theta_\text{p}$) were extracted from the same simulation.

\begin{figure}
    \centering
    \includegraphics{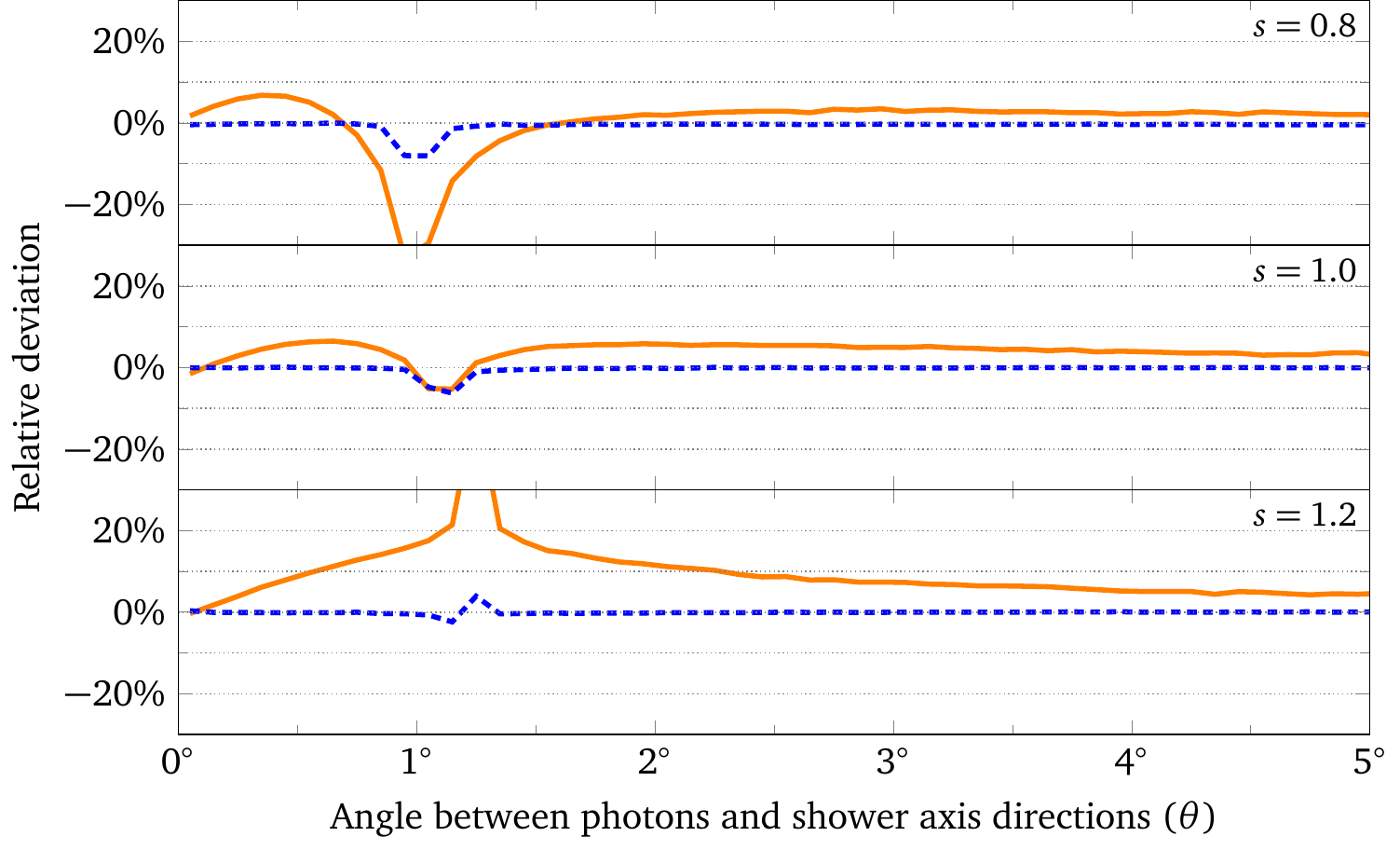}
    \caption{Relative deviation between models of Sections \ref{sec:model:exa} (blue dashed) and \ref{sec:model:app} (orange solid) and the simulated angular distribution of Cherenkov photons in the region of small angles. Curves are shown for three shower ages, indicated in the boxes.}
    \label{fig:model-deviation}
\end{figure}

From Figure \ref{fig:validation} it is seen that in the region of $\theta > 5\degree$ (inset plot) the curves of both models appear superimposed with the reference distributions for the three ages being shown. Further insight about the quality of these models in the region of smaller angles can be obtained by inspection of Figure \ref{fig:model-deviation}, where the relative deviations between both models and the reference distribution are studied.

The exact model presents no deviation with respect to the reference distribution, except in the region around the peak of this distribution where a deviation of $<10\%$ is found, as can be seen in Figure \ref{fig:model-deviation}. However, this may be attributed to a side effect of binning the electron distributions ($\text{d}N_e/\text{d}E$ and $\text{d}N_e/\text{d}\theta_\text{p}$) used as input in Equation \eqref{eq:dedthetadx}.

The approximated model of Section \ref{sec:model:app}, on the other hand, deviates less than $10\%$ from the reference distribution at $s=1.0$ (center plot). At the ages of $s=0.8$ (upper plot) and $s=1.2$ (lower plot), on the other hand, this deviation is typically smaller than $20\%$, except at the peak. While this approximation is not as good as the exact model, it validates the idea that it is possible to approximately reproduce the shape of the angular distribution of Cherenkov photons as a product of two functions, as claimed in Section \ref{sec:model:app}.


\clearpage

\begin{table}
  \centering
  \caption{Coefficients describing the angular distribution of Cherenkov photons in gamma-ray air showers.}
  \begin{tabular}{c c c c c}
    \hline
         $\mu$ & $p_{0,\mu}\,(\pm\mathrm{err})$ & $p_{1,\mu}\,(\pm\mathrm{err})$ & $p_{2,\mu}\,(\pm\mathrm{err})$ & $p_{3,\mu}\,(\pm\mathrm{err})$ \\ 
    \hline
$\nu$      & $  0.34329\,(0.00006)  $ & $ -0.10683\,(0.00002) $  &  $ 1.46852\,(0.00004) $  &                     $-$ \\ 
$\theta_1$ & $   1.4053\,(0.0002)   $ & $  0.32382\,(0.00002) $  &                     $0$  & $-0.048841\,(0.000003)$ \\ 
$\theta_2$ & $  0.95734\,(0.00008)  $ & $  0.26472\,(0.00005) $  &                     $-$  &                     $-$ \\ 
$\epsilon$ & $0.0031206\,(0.0000006)$ & $                   0 $  &                     $0$  &                     $-$ \\ 
    \hline
  \end{tabular}
  \label{tab:par:gamma}     
\end{table}

\begin{table}
  \centering
  \caption{Coefficients describing the angular distribution of Cherenkov photons in proton air showers.}
  \begin{tabular}{c c c c c}
    \hline
         $\mu$ & $p_{0,\mu}\,(\pm\mathrm{err})$ & $p_{1,\mu}\,(\pm\mathrm{err})$ & $p_{2,\mu}\,(\pm\mathrm{err})$ & $p_{3,\mu}\,(\pm\mathrm{err})$ \\ 
    \hline
$\nu$      & $  0.21155\,(0.00006)  $ & $ -0.16639\,(0.00003)  $ & $  1.21803\,(0.00006)  $ & $                    -$ \\ 
$\theta_1$ & $    4.513\,(0.001)    $ & $  0.45092\,(0.00003)  $ & $-0.008843\,(0.000002) $ & $-0.058687 (0.000006) $ \\ 
$\theta_2$ & $  0.90725\,(0.00008)  $ & $  0.41722\,(0.00005)  $ & $                    - $ & $                    -$ \\ 
$\epsilon$ & $ 0.009528\,(0.000002) $ & $ 0.022552\,(0.000007) $ & $  -0.4207\,(0.0002)   $ & $                    -$ \\ 
    \hline
  \end{tabular}
  \label{tab:par:proton}
\end{table}

\clearpage

\end{document}